\documentclass[11pt]{article}           
\usepackage{amsmath}
\usepackage{amssymb}

\newtheorem{Theorem}{Theorem}
\newtheorem{Proposition}{Proposition}

\usepackage{graphics}

\date{}

\title{Binary nonlinearization of the super AKNS system}
\author{
Jingsong He$^{a}$\footnote{Corresponding author, E-mail address:jshe@ustc.edu.cn},
Jing Yu$^{a}$, Ruguang Zhou$^{b}$,Yi Cheng$^{a}$
\vspace{4mm}\\
$^{a}$ Department  of Mathematics, University of Science and
Technology of China,\\ Hefei, Anhei, 230026, P. R. China \\
$^{b}$School of Mathematics Science, Xuzhou Normal University,\\
 Xuzhou, Jiangsu 221116, P. R. China }
\begin{document}
 \maketitle

 \begin{abstract}
We establish the binary nonlinearization approach of the spectral problem of the super AKNS system, and then use it
to obtain the super finite-dimensional integrable Hamiltonian system in supersymmetry manifold $\mathbb{R}^{4N|2N}$.
The super Hamiltonian forms and integrals of motion are given explicitly.


\end{abstract}

\section{Introduction}

The method of nonlinearization of Lax pair for the classical
integrable (1+1)-dimensional system  has aroused strong interests in
soliton theory, including mono-nonlinearization
\cite{cao1,cao2,zhourg1} and binary nonlinearization \cite{ma1,ma2}.
The crucial technique is to find  constraints between the potentials
(i. e, functions of soliton equation) and the eigenfunctions of the
spectral problem associated with the soliton equation. The  soliton
equation can be decomposed into a finite-dimensional integrable
Hamiltonian system (FDIH) by means of above-mentioned constraints.
The first example \cite{cao1} of the nonlinearization for the
(1+1)-dimensional case is the nonlinearization of the (2$\times$ 2)
AKNS system. Moreover, as a natural generalization of this method
for case of (2+1)-dimensional system, the symmetry constraint of the
KP hierarchy is given in references \cite{kon}-\cite{cy1} , which
leads to the intensive research on the constrained KP (cKP)
hierarchy \cite{os1}-\cite{hlc1} based on the pseduo-differential
operator. By this way, a (2+1)-dimensional system can be decomposed
into two (1+1)-dimensional systems. Thus we can understand that
symmetry constraint is one kind of formal variable separation method
for the nonlinear integrable partial differential equations.

On the other hand,  several supersymmetry (susy) integrable systems including supersymmetry AKNS \cite{pop}, super AKNS
system \cite{kup}-\cite{lz3}  and supersymmetry KP hierarchy \cite{mo,lm}, have been studied. Similar to the cKP
hierarchy, the symmetry constraint of the supersymmetry KP hierarchy, i. e.  the constrained supersymmetry KP hierarchy,
is given in reference \cite{ad, anp1}.  Furthermore, there are some  interesting results  on the Lax representation and Hamiltonian
structure of the sAKNS system\cite{ad,ts1},  the "Ghost" symmetry  and the Darboux-Backlund solution of the constrained susy KP\cite{anp1},
super soliton \cite{anp2} and the Hamiltonian structure of the constrained susy KP \cite{ts2}. However, there is no result
on the nonlinearization of the super AKNS (or supersymmetry AKNS) for our best knowledge.  Inspiring  by the relation
between the nonliearization  of the (1+1)-dimensional system and the symmetry constraint of the KP hierarchy,
the appearance of the constrained supersymmetry KP reminds us to ask whether nonlinearization technique can be used in
super AKNS equations and finite dimensional integrable system with fermionic variables can be obtained by this approach.
The purpose of this paper is to give the affirmative answer on the above question for the super AKNS system.

The paper is organized as follows: in  section 2,  we briefly recall
some basic knowledge of the super AKNS hierarchy
\cite{kup}-\cite{lz3}. In section 3, we consider the binary
nonlinearization of spectral problem of the super AKNS system  and
give an explicit constraint. Under this constraint, the super AKNS
system is decomposed into a super FDIH with odd functions. Finally,
we close with some conclusions and discussions in section 4.

\section{Super AKNS Hierarchy}
Let us start with the following spectral problem\cite{mo,Lz1+2}
\begin{equation}\label{1}
\phi_x=M\phi, \quad  M=\left(\begin{array}{ccc}
-\lambda&q&\alpha\\
r&\lambda&\beta\\
\beta&-\alpha&0\\
 \end{array}\right),\quad
 \phi= \left(\begin{array}{ccc}
\phi_1\\
\phi_2\\
\phi_3
 \end{array}\right),\end{equation}
where $\lambda$, $q$ and  $r$ are even elements such that $p(\lambda)=p(q)=p(r)=0$; but  $\alpha$ and
$\beta$ are odd elements such that $p(\alpha)=p(\beta)=1$.  $\lambda$ is a eigenparameter of this system.
$q, r, \alpha$ and  $\beta$ are functions of  the usual time-space variables $x$ and $ t$. Here small $p(f)$ denotes the
parity of the arbitrary function $f$. Note that the main reference of supersymmetry and susy analysis used in this paper  is
literature \cite{cartier}. Note that $M \in B(0,1)$, $B(0,1)$ is a  Lie superalegra.

In order to obtain the super AKNS hierarchy, we first solve
co-adjoint equation associated with (\ref{1})
\begin{equation}\label{2}
N_x=[M, N]=MN-NM,
\end{equation}
with$$N=\left(\begin{array}{ccc}
A&B&\rho\\
C&-A&\delta\\
\delta&-\rho&0
\end{array}\right)
=\sum\limits_{j=0}^{\infty}\left(\begin{array}{ccc}
a_j&b_j&\rho_j\\
c_j&-a_j&\delta_j\\
\delta_j&-\rho_j&0
\end{array}\right)\lambda^{-j},$$
where p(A)=p(B)=p(C)=0, p($\rho$)=p($\delta$)=1 and $a_j, b_j, c_j,
\rho_j, \delta_j (j\geq0)$ are determined later.

Substituting  M, N into Eq. (\ref{2}) and comparing the coefficients of $\lambda^{-j} (j\geq 0)$, we have
\begin{eqnarray}\label{3}\left\{\begin{array}{l}
b_0=c_0=\rho_0=\delta_0=0,\\
a_{j, x}=qc_j-rb_j+\alpha\delta_j+\beta\rho_j,\qquad j\geq 0,\\
b_{j, x}=-2b_{j+1}-2qa_j-2\alpha\rho_j, \qquad j\geq 0,\\
c_{j, x}=2c_{j+1}+2ra_j+2\beta\delta_j, \qquad j\geq 0,\\
\rho_{j, x}=-\rho_{j+1}+q\delta_j-\alpha a_j-\beta b_j,\qquad j\geq 0,\\
\delta_{j, x}=\delta_{j+1}+r\rho_j-\alpha c_j+\beta a_j,\qquad j\geq
0.
\end{array}\right.\end{eqnarray}
Equation (\ref{3}) can also be rewritten as
\begin{eqnarray}\label{4}
(c_{j+1}, b_{j+1}, \delta_{j+1}, \rho_{j+1})^{T} =L(c_j, b_j,
\delta_j, \rho_j)^{T},\end{eqnarray}where
\begin{eqnarray}L=\left(\begin{array}{cccc}
\frac{1}{2}\partial_x-r\partial_x^{-1}q&r\partial_x^{-1}r&-\beta-r\partial_x^{-1}\alpha&-r\partial_x^{-1}\beta\\
-q\partial_x^{-1}q&-\frac{1}{2}\partial_x+q\partial_x^{-1}r&-q\partial_x^{-1}\alpha&-\alpha-q\partial_x^{-1}\beta\\
\alpha-\beta\partial_x^{-1}q&\beta\partial_x^{-1}r&\partial_x-\beta\partial_x^{-1}\alpha&-r-\beta\partial_x^{-1}\beta\\
-\alpha\partial_x^{-1}q&-\beta+\alpha\partial_x^{-1}r&q-\alpha\partial_x^{-1}\alpha&-\partial_x-\alpha\partial_x^{-1}\beta
\end{array}\right).\end{eqnarray}
For a given initial value, the  $a_j, b_j, c_j, \rho_j, \delta_j (j\geq 1)$ can be calculated by the recursion relation
(\ref{4}). In particular,  let $a_0=-1$, we have
\begin{eqnarray*}a_1=0, b_1=q, c_1=r, \rho_1=\alpha, \delta_1=\beta,\\
a_2=\frac{1}{2}q r+\alpha\beta, b_2=-\frac{1}{2}q_x,
c_2=\frac{1}{2}r_x, \rho_2=-\alpha_x,
\delta_2=\beta_x.\end{eqnarray*}

Then, let us consider the spectral problem (\ref{1}) with the
following auxiliary spectral problem
\begin{equation}\label{5}
\phi_{t_n}=N^{(n)}\phi=(\lambda^nN)_{+}\phi,
\end{equation}
with
$$(\lambda^nN)_{+}=\sum\limits_{j=0}^{n}\left(\begin{array}{ccc}
a_j&b_j&\rho_j\\
c_j&-a_j&\delta_j\\
\delta_j&-\rho_j&0\end{array}\right)\lambda^{n-j},$$ where the
symbol "+" denotes taking the nonnegative power of $\lambda$.
A simple calculation leads to $$N^{(1)}=\left(\begin{array}{ccc}
-\lambda&q&\alpha\\
r&\lambda&\beta\\
\beta&-\alpha&0\end{array}\right)=M,$$
\begin{eqnarray}
N^{(2)}=\left(\begin{array}{ccc}
-\lambda^{2}+\frac{1}{2}qr+\alpha\beta&q\lambda-\frac{1}{2}q_x&\alpha\lambda-\alpha_x\\
r\lambda+\frac{1}{2}r_x&\lambda^{2}-\frac{1}{2}qr-\alpha\beta&\beta\lambda+\beta_x\\
\beta\lambda+\beta_x&-\alpha\lambda+\alpha_x&0\end{array}\right).
\end{eqnarray}
From the compatible condition $\phi_{x, t_n}=\phi_{t_n, x}$ according to equations (\ref{1}) and (\ref{5}),
we get a zero curvature equation
\begin{equation}\label{6}
M_{t_n}-N^{(n)}_x+[M, N^{(n)}]=0,
\end{equation}
which gives the super AKNS hierarchy
\begin{eqnarray}\label{7}\left\{\begin{array}{l}
q_{t_n}=b_{n, x}+2qa_n+2\alpha\rho_n=-2b_{n+1},\\
r_{t_n}=c_{n, x}-2ra_n-2\beta\delta_n=2c_{n+1},\\
\alpha_{t_n}=\rho_{n, x}-q\delta_n+\alpha a_n+\beta b_n=-\rho_{n+1},\\
\beta_{t_n}=\delta_{n, x}-\beta a_n+\alpha c_n-r\rho_n=\delta_{n+1}.
\end{array}\right.\end{eqnarray}
Setting $n=2$, eq.(\ref{7}) gives
\begin{equation}\left\{\begin{array}{l}
q_{t_2}=-\frac{1}{2}q_{xx}+q^{2}r+2q\alpha\beta-2\alpha\alpha_x,\\
r_{t_2}=\frac{1}{2}r_{xx}-qr^{2}-2r\alpha\beta-2\beta\beta_x,\\
\alpha_{t_2}=-\alpha_{xx}-q\beta_x+\frac{1}{2}qr\alpha-\frac{1}{2}q_x\beta,\\
\beta_{t_2}=\beta_{xx}+r\alpha_x+\frac{1}{2}r_x\alpha-\frac{1}{2}qr\beta.
\end{array}\right.\end{equation}
Setting  $n=3$  in the super AKNS hierarchy in eq(\ref{7}) results in
\begin{equation}\left\{\begin{array}{l}
q_{t_3}=\frac{1}{4}q_{xxx}-\frac{3}{2}qq_xr-3q\alpha_x\beta+3q\alpha\beta_x+3\alpha\alpha_{xx},\\
r_{t_3}=\frac{1}{4}r_{xxx}-\frac{3}{2}qrr_x+3r\alpha_x\beta-3r\alpha\beta_x-3\beta\beta_{xx},\\
\alpha_{t_3}=\alpha_{xxx}+\frac{3}{2}q_x\beta_x-\frac{3}{4}q_xr\alpha-\frac{3}{4}qr_x\alpha
-\frac{3}{2}qr\alpha_x+\frac{3}{4}q_{xx}\beta,\\
\beta_{t_3}=\beta_{xxx}+\frac{3}{2}r_x\alpha_x+\frac{3}{4}r_{xx}\alpha-\frac{3}{4}q_xr\beta
-\frac{3}{4}qr_x\beta-\frac{3}{2}qr\beta_x.
\end{array}\right.\end{equation}

 Introducing vectors
$U_0=(r, q, \beta, \alpha)^{T}$, $U=(r, -q, 2\beta, -2\alpha)^{T}$,
(\ref{7}) becomes a compact form
\begin{equation}\label{8}
U_{t_n}=\left(\begin{array}{cccc}
r\\-q\\2\beta\\-2\alpha\end{array}\right)_{t_n}
=2\left(\begin{array}{cccc}c_{n+1}\\ b_{n+1}\\ \delta_{n+1}\\
\rho_{n+1}\end{array}\right)=2L^{n}U_0.\end{equation} Let
\begin{equation}\label{L1}L_1=\left(\begin{array}{cccc}
-\frac{1}{2}\partial_x+q\partial_x^{-1}r&-q\partial_x^{-1}q&
\frac{1}{2}\alpha+\frac{1}{2}q\partial_x^{-1}\beta&-\frac{1}{2}q\partial_x^{-1}\alpha\\
r\partial_x^{-1}r&\frac{1}{2}\partial_x-r\partial_x^{-1}q&\frac{1}{2}r\partial_x^{-1}\beta
&-\frac{1}{2}\beta-\frac{1}{2}r\partial_x^{-1}\alpha\\
2\beta-2\alpha\partial_x^{-1}r&2\alpha\partial_x^{-1}q&-\partial_x-\alpha\partial_x^{-1}\beta&
-q+\alpha\partial_x^{-1}\alpha\\
2\beta\partial_x^{-1}r&2\alpha-2\beta\partial_x^{-1}q&r+\beta\partial_x^{-1}\beta&
\partial_x-\beta\partial_x^{-1}\alpha\end{array}\right),\end{equation}
and
$$\tilde{U}=(r, q, \beta, \alpha)^{T}=\frac{1}{2}
\left(\begin{array}{cccc}
2&0&0&0\\0&-2&0&0\\0&0&1&0\\0&0&0&-1\end{array}\right)\left(\begin{array}{cccc}
r\\-q\\2\beta\\-2\alpha\end{array}\right)=\frac{1}{2}gU,$$
$$\tilde{U}_0=\left(\begin{array}{cccc}q\\r\\-2\alpha\\2\beta\end{array}\right)
=J^{-1}g\tilde{U},\quad
g=\left(\begin{array}{cccc}2&0&0&0\\0&-2&0&0\\0&0&1&0\\0&0&0&-1\end{array}\right),$$
\begin{equation}\label{J}J=\left(\begin{array}{cccc}0&2&0&0\\
-2&0&0&0\\0&0&0&\frac{1}{2}\\0&0&\frac{1}{2}&0\end{array}\right).\end{equation}

It is easy to verify
\begin{equation}\label{9*}L_1=J^{-1}gLg^{-1}J,\end{equation}
thus
$$L_1^{n}=J^{-1}gL^{n}g^{-1}J,$$
and Eq. (\ref{8}) becomes
\begin{equation}\label{9}\tilde{U}_{t_n}=JL_1^{n}\tilde{U}_0.\end{equation}
Using the supertrace identity \cite{Hu}
\begin{equation}\label{11}
\frac{\delta}{\delta U_0}\int Str(N\frac{\partial
M}{\partial\lambda})dx=(\lambda^{-\gamma}(\frac{\partial}{\partial\lambda})\lambda^{\gamma})Str(\frac{\partial
M}{\partial U_0}N),\end{equation}  then
\begin{equation}\label{12}\left(\begin{array}{cccc}
\frac{\delta}{\delta r}\\\frac{\delta}{\delta
q}\\\frac{\delta}{\delta\beta}\\\frac{\delta}{\delta\alpha}\end{array}\right)(-2a_{n+2})=(\gamma-n-1)
\left(\begin{array}{cccc}b_{n+1}\\c_{n+1}\\-2\rho_{n+1}\\2\delta_{n+1}\end{array}\right).\end{equation}
Let $n=0$, then $\gamma=0$.

Moreover, the  equation(\ref{9}) can be written as the following super Hamiltonian form\cite{lz3}
\begin{equation}\label{13}
\tilde{U}_{t_n}=JL_1^{n}\left(\begin{array}{cccc}
q\\r\\-2\alpha\\2\beta\end{array}\right)=J\frac{\delta H_n}{\delta
U_0},\qquad\frac{\delta H_n}{\delta
U_0}=\left(\begin{array}{cccc}b_{n+1}\\c_{n+1}\\-2\rho_{n+1}\\2\delta_{n+1}\end{array}\right),\qquad
H_n=\int\frac{2}{n+1}a_{n+2}dx,\end{equation} where J is the supersymplectic operator.
We would like to point out that the authors in reference \cite{lz3} calculated the Hamiltonian form by a direct constrained
variation method instead of supertrace identity.

\section{Binary Nonlinearization}
Now we are in a position to discuss the nonlinearization of the super AKNS hierarchy given in the last section.
To this aim, consider the spectral problem in eq.(\ref{1}) and its adjoint spectral problem
\begin{equation}\label{14}
\psi_x=-M^{St}\psi=\left(\begin{array}{ccc}\lambda&-r&\beta\\-q&-\lambda&-\alpha\\-\alpha&-\beta&0\end{array}\right)\psi,
\quad
\psi=\left(\begin{array}{ccc}\psi_1\\\psi_2\\\psi_3\end{array}\right),\end{equation}
where St means supertranspose \cite{cartier}.

It is not difficult to get
\begin{eqnarray}\label{16}
\frac{\delta\lambda}{\delta U_0}=\left(\begin{array}{c}
\frac{\delta\lambda}{\delta q}\\
\frac{\delta\lambda}{\delta r}\\
\frac{\delta\lambda}{\delta \alpha}\\
\frac{\delta\lambda}{\delta \beta}\end{array}\right)=
\left(\begin{array}{c}\psi_1\phi_2\\
\psi_2\phi_1\\
\psi_1\phi_3+\psi_3\phi_2\\
\psi_2\phi_3-\psi_3\phi_1\end{array}\right).
\end{eqnarray}
by a similar way of the counterpart of in the AKNS system\cite{ma1,ma2} .
When zero boundary conditions
 $\lim_{|x|\rightarrow\infty}\phi=\lim_{|x|\rightarrow\infty}\psi=0$ are
 imposed, we can verify a simple characteristic property of the
 variational derivative of $\lambda$
\begin{equation}\label{14*}
L_1\frac{\delta\lambda}{\delta
U_0}=\lambda\frac{\delta\lambda}{\delta U_0},\end{equation} where
$L_1$ is defined by (\ref{L1}).

 Choosing N distinct eigenparameters
$\lambda_1,\cdots,\lambda_N$, equation (\ref{13}) becomes
\begin{equation}\label{22}\left(\begin{array}{cccc}
\frac{\delta}{\delta r}\\\frac{\delta}{\delta q}\\
\frac{\delta}{\delta\beta}\\\frac{\delta}{\delta\alpha}\end{array}\right)H_k
=\left(\begin{array}{cccc}b_{k+1}\\c_{k+1}\\-2\rho_{k+1}\\2\delta_{k+1}\end{array}\right)
=\sum\limits_{j=1}^{N}\left(\begin{array}{cccc}
\frac{\delta\lambda_j}{\delta r}\\
\frac{\delta\lambda_j}{\delta q}\\
\frac{\delta\lambda_j}{\delta\beta}\\
\frac{\delta\lambda_j}{\delta\alpha}\end{array}\right)=
\left(\begin{array}{cccc}<\Psi_2, \Phi_1>\\<\Psi_1, \Phi_2>\\
<\Psi_2, \Phi_3>-<\Psi_3, \Phi_1>\\ <\Psi_1, \Phi_3>+<\Psi_3,
\Phi_2>\end{array}\right),\end{equation} where $\Phi_i=(\phi_{i1},
\cdots, \phi_{iN})^{T}, \Psi_i=(\psi_{i1}, \cdots, \psi_{iN})^{T},$
i=1, 2, 3, and $<.,.>$ denotes the inner product in $R^{N}$.

Let k=0 in (\ref{22}), then it gives
\begin{equation}\label{15}\left\{\begin{array}{l}
q=<\Psi_2, \Phi_1>,\\
r=<\Psi_1, \Phi_2>,\\
\alpha=-\frac{1}{2}(<\Psi_2, \Phi_3>-<\Psi_3,
\Phi_1>),\\
\beta=\frac{1}{2}(<\Psi_1, \Phi_3>+<\Psi_3,
\Phi_2>).\end{array}\right.\end{equation} Substituting (\ref{15})
into the spectral problem(\ref{1}) and the adjoint spectral
problem(\ref{14}), we obtain the following finite-dimensional
systems
\begin{equation}\label{16}\left\{\begin{array}{l}
\phi_{1j, x}=-\lambda_j\phi_{1j}+<\Psi_2,
\Phi_1>\phi_{2j}-\frac{1}{2}(<\Psi_2, \Phi_3>-<\Psi_3,
\Phi_1>)\phi_{3j},\quad 1\leq j\leq N,\\
\phi_{2j, x}=<\Psi_1,
\Phi_2>\phi_{1j}+\lambda_j\phi_{2j}+\frac{1}{2}(<\Psi_1,
\Phi_3>+<\Psi_3,
\Phi_2>)\phi_{3j},\quad 1\leq j\leq N,\\
\phi_{3j, x}=\frac{1}{2}(<\Psi_1, \Phi_3>+<\Psi_3,
\Phi_2>)\phi_{1j}+\frac{1}{2}(<\Psi_2, \Phi_3>-<\Psi_3, \Phi_1>)\phi_{2j},
\quad 1\leq j\leq N,\\
\psi_{1j, x}=\lambda_j\psi_{1j}-<\Psi_1,
\Phi_2>\psi_{2j}+\frac{1}{2}(<\Psi_1, \Phi_3>+<\Psi_3, \Phi_2>)\psi_{3j},
\quad 1\leq j\leq N,\\
\psi_{2j, x}=-<\Psi_2,
\Phi_1>\psi_{1j}-\lambda_j\psi_{2j}+\frac{1}{2}(<\Psi_2,
\Phi_3>-<\Psi_3,
\Phi_1>)\psi_{3j},\quad 1\leq j\leq N,\\
\psi_{3j, x}=\frac{1}{2}(<\Psi_2, \Phi_3>-<\Psi_3,
\Phi_1>)\psi_{1j}-\frac{1}{2}(<\Psi_1, \Phi_3>+<\Psi_3,
\Phi_2>)\psi_{2j},\quad 1\leq j\leq
N,\end{array}\right.\end{equation} which can  be written as the
following super Hamiltonian form
\begin{equation}\left\{\begin{array}{l}
\Phi_{1, x}=\frac{\partial H_1}{\partial \Psi_1}, \Phi_{2,
x}=\frac{\partial H_1}{\partial \Psi_2},
\Phi_{3, x}=\frac{\partial H_1}{\partial \Psi_3},\\
\Psi_{1, x}=-\frac{\partial H_1}{\partial \Phi_1}, \Psi_{2,
x}=-\frac{\partial H_1}{\partial \Phi_2}, \Psi_{3, x}=\frac{\partial
H_1}{\partial \Phi_3},\end{array}\right.\end{equation} where
\begin{eqnarray*}
H_1&=&-<\Lambda \Psi_1, \Phi_1>+<\Lambda \Psi_2, \Phi_2>+<\Psi_2,
\Phi_1><\Psi_1, \Phi_2>\\&&-\frac{1}{2}(<\Psi_2, \Phi_3>-<\Psi_3,
\Phi_1>)(<\Psi_1, \Phi_3>+<\Psi_3, \Phi_2>).\end{eqnarray*}

As for $t_2$-part, we consider the following spectral problem
\begin{equation}\label{19}
\phi_{t_2}=N^{(2)}\phi=\left(\begin{array}{ccc}
-\lambda^{2}+\frac{1}{2}qr+\alpha\beta&q\lambda-\frac{1}{2}q_x&\alpha\lambda-\alpha_x\\
r\lambda+\frac{1}{2}r_x&\lambda^{2}-\frac{1}{2}qr-\alpha\beta&\beta\lambda+\beta_x\\
\beta\lambda+\beta_x&-\alpha\lambda+\alpha_x&0\end{array}\right)\phi,\end{equation}
and its adjoint spectral problem
\begin{equation}\label{20}
\psi_{t_2}=-(N^{(2)})^{St}\psi=\left(\begin{array}{ccc}
\lambda^{2}-\frac{1}{2}qr-\alpha\beta&-r\lambda-\frac{1}{2}r_x&\beta\lambda+\beta_x\\
-q\lambda+\frac{1}{2}q_x&-\lambda^{2}+\frac{1}{2}qr+\alpha\beta&-\alpha\lambda+\alpha_x\\
-\alpha\lambda+\alpha_x&-\beta\lambda-\beta_x&0\end{array}\right)\psi.\end{equation}
Substituting equation (\ref{15}) into spectral problems (\ref{19}),
(\ref{20}) and noticing systems (\ref{16}), we obtain the following
finite-dimensional system
\begin{equation}\label{21}\left\{\begin{array}{l}
\phi_{1j, t_2}=(-\lambda_j^{2}+\frac{1}{2}\tilde{q}\tilde{r}+\tilde{\alpha}\tilde{\beta})\phi_{1j}
+(\tilde{q}\lambda_j-\frac{1}{2}\tilde{q}_x)\phi_{2j}+(\tilde{\alpha}\lambda_j-\tilde{\alpha}_x)
\phi_{3j},\quad 1\leq j\leq N,\\
\phi_{2j,
t_2}=(\tilde{r}\lambda_j+\frac{1}{2}\tilde{r}_x)\phi_{1j}+(\lambda_j^{2}-
\frac{1}{2}\tilde{q}\tilde{r}-\tilde{\alpha}\tilde{\beta})\phi_{2j}
+(\tilde{\beta}\lambda_j+\tilde{\beta}_x)\phi_{3j},\quad 1\leq j\leq N,\\
\phi_{3j, t_2}=(\tilde{\beta}\lambda_j+\tilde{\beta}_x)\phi_{1j}
+(-\tilde{\alpha}\lambda_j+\tilde{\alpha}_x)\phi_{2j},\quad 1\leq j\leq N,\\
\psi_{1j,
t_2}=(\lambda_j^{2}-\frac{1}{2}\tilde{q}\tilde{r}-\tilde{\alpha}\tilde{\beta})\psi_{1j}
-(\tilde{r}\lambda_j+\frac{1}{2}\tilde{r}_x)\psi_{2j}+(\tilde{\beta}\lambda_j+
\tilde{\beta}_x)\psi_{3j},\quad 1\leq j\leq N,\\
\psi_{2j, t_2}=-(\tilde{q}\lambda_j-\frac{1}{2}\tilde{q}_x)\psi_{1j}
+(-\lambda_j^{2}+\frac{1}{2}\tilde{q}\tilde{r}+\tilde{\alpha}\tilde{\beta})\psi_{2j}
+(-\tilde{\alpha}\lambda_j+\tilde{\alpha}_x)\psi_{3j},\quad 1\leq j\leq N,\\
\psi_{3j,
t_2}=(-\tilde{\alpha}\lambda_j+\tilde{\alpha}_x)\psi_{1j}-(\tilde{\beta}\lambda_j+
\tilde{\beta}_x)\psi_{2j},\quad 1\leq j\leq
N,\end{array}\right.\end{equation} where $\tilde{q}, \tilde{r},
\tilde{\alpha}, \tilde{\beta}$ denote the potentials under the
constraint (\ref{15}).
Here $\tilde{P}$ denote the new expression generated from $P(u)$ by the
the binary nonlinear constraint (\ref{15}).

By a direct but tedious calculation,  the finite system in eq.(\ref{21}) can be
written as the following super Hamiltonian form
\begin{equation}\left\{\begin{array}{l} \Phi_{1,
t_2}=\frac{\partial H_2}{\partial\Psi_1}, \Phi_{2,
t_2}=\frac{\partial H_2}{\partial\Psi_2},
\Phi_{3, t_2}=\frac{\partial H_2}{\partial\Psi_3},\\
\Psi_{1, t_2}=-\frac{\partial H_2}{\partial\Phi_1}, \Psi_{2,
t_2}=-\frac{\partial H_2}{\partial\Phi_2}, \Psi_{3,
t_2}=\frac{\partial H_2}{\partial
\Phi_3},\end{array}\right.\end{equation} where
\begin{eqnarray*}
H_2&=&<\Lambda^2\Psi_2, \Phi_2>-<\Lambda^{2}\Psi_1, \Phi_1>+<\Psi_2,
\Phi_1><\Lambda \Psi_1, \Phi_2> +<\Lambda \Psi_2, \Phi_1><\Psi_1,
\Phi_2>\\&&-\frac{1}{4}(<\Psi_1, \Phi_1>-<\Psi_2, \Phi_2>)(<\Psi_2,
\Phi_3>-<\Psi_3, \Phi_1>)(<\Psi_1, \Phi_3>+<\Psi_3, \Phi_2>)\\
&& +\frac{1}{2}<\Psi_2, \Phi_1><\Psi_1, \Phi_2>(<\Psi_1,
\Phi_1>-<\Psi_2, \Phi_2>)-\frac{1}{2}(<\Psi_2, \Phi_3>-<\Psi_3,
\Phi_1>)\\
&&(<\Lambda \Psi_1, \Phi_3> +<\Lambda \Psi_3,
\Phi_2>)-\frac{1}{2}(<\Lambda \Psi_2, \Phi_3>-<\Lambda \Psi_3,
\Phi_1>)(<\Psi_1, \Phi_3>+<\Psi_3, \Phi_2>) .\end{eqnarray*}

Making use of (\ref{14*}), the recursion relation (\ref{4}) and
equation(\ref{9*}), we have
\begin{equation}\left\{\begin{array}{l}
\tilde{a}_i=\frac{1}{2}<\Lambda^{i-1}\Psi_1, \Phi_1>-\frac{1}{2}<\Lambda^{i-1}\Psi_2, \Phi_2>,
\quad i\geq1,\\
\tilde{b}_i=<\Lambda^{i-1}\Psi_2, \Phi_1>,\quad i\geq1,\\
\tilde{c}_i=<\Lambda^{i-1}\Psi_1, \Phi_2>,\quad i\geq1,\\
\tilde{\rho}_i=-\frac{1}{2}(<\Lambda^{i-1}\Psi_2, \Phi_3>-<\Lambda^{i-1}\Psi_3, \Phi_1>),
\quad i\geq1,\\
\tilde{\delta}_i=\frac{1}{2}(<\Lambda^{i-1}\Psi_1,
\Phi_3>+<\Lambda^{i-1}\Psi_3, \Phi_2>),\quad i\geq1.
\end{array}\right.\end{equation}
Then $\tilde{N}_x=[\tilde{M}, \tilde{N}]$ is still satisfied, and
$(\tilde{N}^{2})_x= [\tilde{M}, \tilde{N}^{2}]$ is also satisfied.
Therefore
$$F_x=(\frac{1}{2}Str\tilde{N}^{2})_x=\frac{1}{2}Str(\tilde{N}^{2})_x=\frac{d}{dx}
(\tilde{a}^{2}+\tilde{b}\tilde{c}+2\tilde{\rho}\tilde{\delta})=0.$$
The identity indicates that F is a generating function of integrals
of motion for the nonlinearized spatial systems (\ref{16}). Let
$F=\sum\limits_{n\geq0}F_n\lambda^{-n}$, we obtain the following
formulas
 \begin{eqnarray}F_1=-2\tilde{a}_1=-<\Psi_1,
\Phi_1>+<\Psi_2, \Phi_2>,\end{eqnarray}
\begin{eqnarray}
F_n&=&\sum\limits_{i=1}^{n-1}(\tilde{a}_i\tilde{a}_{n-i}+\tilde{b}_i\tilde{c}_{n-i}+2\tilde{\rho}_i\tilde{\delta}_{n-i})
+2\tilde{a}_0\tilde{a}_n\nonumber\\&=&
\sum\limits_{i=1}^{n-1}[\frac{1}{4}(<\Lambda^{i-1}\Psi_1,
\Phi_1>-<\Lambda^{i-1}\Psi_2, \Phi_2>)(<\Lambda^{n-i-1}\Psi_1,
\Phi_1>-<\Lambda^{n-i-1}\Psi_2,
\Phi_2>)\nonumber\\&&-\frac{1}{2}(<\Lambda^{i-1}\Psi_2,
\Phi_3>-<\Lambda^{i-1}\Psi_3, \Phi_1>)(<\Lambda^{n-i-1}\Psi_1,
\Phi_3>+<\Lambda^{n-i-1}\Psi_3,
\Phi_2>)\nonumber\\&&+<\Lambda^{i-1}\Psi_2,
\Phi_1><\Lambda^{n-i-1}\Psi_1, \Phi_2>]-<\Lambda^{n-1}\Psi_1,
\Phi_1>+<\Lambda^{n-1}\Psi_2, \Phi_2>.\end{eqnarray}

Let us consider the temporal part of the super AKNS hierarchy in eq.(\ref{5})
\begin{equation}\label{34}\left\{\begin{array}{l}\left(\begin{array}{c}
\phi_{1j}\\\phi_{2j}\\\phi_{3j}\end{array}\right)_{t_n}
=\left(\begin{array}{ccc}
\sum_{i=0}^{n}\tilde{a}_i\lambda_j^{n-i}&\sum_{i=0}^{n}\tilde{b}_i\lambda_j^{n-i}
&\sum_{i=0}^{n}\tilde{\rho}_i\lambda_j^{n-i}\\\sum_{i=0}^{n}\tilde{c}_i\lambda_j^{n-i}
&-\sum_{i=0}^{n}\tilde{a}_i\lambda_j^{n-i}&\sum_{i=0}^{n}\tilde{\delta}_i\lambda_j^{n-i}\\
\sum_{i=0}^{n}\tilde{\delta}_i\lambda_j^{n-i}&-\sum_{i=0}^{n}\tilde{\rho}_i\lambda_j^{n-i}&0
\end{array}\right)\left(\begin{array}{c}\phi_{1j}\\\phi_{2j}\\\phi_{3j}\end{array}\right)
,\quad 1\leq j\leq N,\\
\left(\begin{array}{c}
\psi_{1j}\\\psi_{2j}\\\psi_{3j}\end{array}\right)_{t_n}
=\left(\begin{array}{ccc}
-\sum_{i=0}^{n}\tilde{a}_i\lambda_j^{n-i}&-\sum_{i=0}^{n}\tilde{c}_i\lambda_j^{n-i}
&\sum_{i=0}^{n}\tilde{\delta}_i\lambda_j^{n-i}\\-\sum_{i=0}^{n}\tilde{b}_i\lambda_j^{n-i}
&\sum_{i=0}^{n}\tilde{a}_i\lambda_j^{n-i}&-\sum_{i=0}^{n}\tilde{\rho}_i\lambda_j^{n-i}\\
-\sum_{i=0}^{n}\tilde{\rho}_i\lambda_j^{n-i}&-\sum_{i=0}^{n}\tilde{\delta}_i\lambda_j^{n-i}&0
\end{array}\right)\left(\begin{array}{c}\psi_{1j}\\\psi_{2j}\\\psi_{3j}\end{array}\right)
,\quad 1\leq j\leq N.\end{array}\right.\end{equation}
 A direct calculation leads to
\begin{equation}\left\{\begin{array}{l}
\Phi_{1, t_n}=\frac{\partial F_{n+1}}{\partial\Psi_1},\quad \Phi_{2,
t_n}=\frac{\partial F_{n+1}}{\partial\Psi_2},\quad \Phi_{3,
t_n}=\frac{\partial F_{n+1}}{\partial\Psi_3},\\
\Psi_{1, t_n}=-\frac{\partial F_{n+1}}{\partial\Phi_1},\quad
\Psi_{2, t_n}=-\frac{\partial F_{n+1}}{\partial\Phi_2},\quad
\Psi_{3, t_n}=\frac{\partial F_{n+1}}{\partial\Phi_3}.
\end{array}\right.\end{equation}
For example,
\begin{eqnarray*}
\Phi_{1, t_n}&=&\sum_{i=0}^{n}\tilde{a}_i\Lambda^{n-i}\Phi_1
+\sum_{i=1}^{n}\tilde{b}_i\Lambda^{n-i}\Phi_2
+\sum_{i=1}^{n}\tilde{\rho}_i\Lambda^{n-i}\Phi_3\\
&=&-\Lambda^{n}\Phi_1+\sum_{i=1}^{n}[\frac{1}{2}(<\Lambda^{i-1}\Psi_1,
\Phi_1>-<\Lambda^{i-1}\Psi_2,
\Phi_2>)\Lambda^{n-i}\Phi_1+<\Lambda^{i-1}\Psi_2,
\Phi_1>\Lambda^{n-i}\Phi_2\\&&-\frac{1}{2}(<\Lambda^{i-1}\Psi_2,
\Phi_3>-<\Lambda^{i-1}\Psi_3, \Phi_1>)\Lambda^{n-i}\Phi_3]\\
&=&\frac{\partial F_{n+1}}{\partial\Psi_1}.\end{eqnarray*}

 It is not difficult to see that
$F_n (n\geq0)$ are  also  integrals of motion for
equation(\ref{34}), i. e.
$$\{ F_{m+1}, F_{n+1}\}=\frac{\partial}{\partial t_n}F_{m+1}=0, m,
n\geq0,$$ where Poisson bracket is defined by
\begin{equation}\{ F, G\}=\sum_{i=1}^{3}\sum_{j=1}^{N}
(\frac{\partial F}{\partial\phi_{ij}}\frac{\partial
G}{\partial\psi_{ij}}-(-1)^{p(\phi_{ij})p(\psi_{ij})}\frac{\partial
F}{\partial\psi_{ij}}\frac{\partial
G}{\partial\phi_{ij}}).\end{equation}

With the help of the result of nonlinearization\cite{ma1,ma2,ma3},
it is natural for us to set
\begin{equation}
f_k=\psi_{1k}\phi_{1k}+\psi_{2k}\phi_{2k}+\psi_{3k}\phi_{3k},\quad
1\leq k\leq N,
\end{equation}
 and  verify they are also integrals of motion of the
constrained systems (\ref{16}) and (\ref{34}). In what follows, we
will give a proposition to show the independence of
$\{f_k\}_{k=1}^{N}$, $\{F_m\}_{m=1}^{2N}$.

\begin{Proposition}\label{propind} The integrals of motion
$\{f_k\}_{k=1}^{N}$ and $\{F_m\}_{m=1}^{2N}$ are functionally
independent over some region of the supersymmetry manifold  $\mathbb{R}^{4N|2N}$. Here the definition of $
\mathbb{R}^{M|N}$\cite{cartier}
is given by $(x^1, \cdots, x^M,\xi^1, \cdots,\xi^N)$ with $x^i\in \mathbb{R}$ while $\xi^i$ are odd varibales.
\end{Proposition}
\noindent{\bf Proof}\ \  Suppose that the result of the proposition is not true, that
is to say, there does not exist any region of $\mathbb{R}^{4N|2N}$ over which
the integrals of motion $\{f_k\}_{k=1}^{N}$, $\{F_m\}_{m=1}^{2N}$
are functionally independent. Therefore, there exist 3N constants
$\{\alpha_k\}_{k=1}^{N}$, $\{\beta_m\}_{m=1}^{2N}$ which do not
equal to zero at the same time, so that we have for all points in
$\mathbb{R}^{4N|2N}$
\begin{equation}\label{17}
\sum_{k=1}^{N}\alpha_k((\frac{\partial f_k}{\partial\Phi_1})^{T},
(\frac{\partial f_k}{\partial\Phi_2})^{T}, (\frac{\partial
f_k}{\partial\Phi_3})^{T})+\sum_{m=1}^{2N}\beta_m((\frac{\partial
F_m}{\partial\Phi_1})^{T}, (\frac{\partial
F_m}{\partial\Phi_2})^{T}, (\frac{\partial
F_m}{\partial\Phi_3})^{T})=0.\end{equation} After a direct
calculation, we have
\begin{eqnarray*}\left\{\begin{array}{l}
\frac{\partial
f_k}{\partial\phi_{1j}}|_{\Phi_1=\Phi_2=0}=\delta_{jk}\psi_{1j},\quad
1\leq k, j\leq N, \\
\frac{\partial
f_k}{\partial\phi_{2j}}|_{\Phi_1=\Phi_2=0}=\delta_{jk}\psi_{2j},\quad
1\leq k, j\leq N, \\
\frac{\partial
f_k}{\partial\phi_{3j}}|_{\Phi_1=\Phi_2=0}=-\delta_{jk}\psi_{3j},\quad
1\leq k, j\leq N, \\
\frac{\partial
F_m}{\partial\Phi_1}|_{\Phi_1=\Phi_2=0}=-\frac{1}{2}\sum_{i=1}^{m-1}
<\Lambda^{m-i-1}\Psi_1,
\Phi_3>\Lambda^{i-1}\Psi_3-\Lambda^{m-1}\Psi_1,\quad
1\leq m\leq 2N,\\
\frac{\partial
F_m}{\partial\Phi_2}|_{\Phi_1=\Phi_2=0}=-\frac{1}{2}\sum_{i=1}^{m-1}
<\Lambda^{i-1}\Psi_2,
\Phi_3>\Lambda^{m-i-1}\Psi_3+\Lambda^{m-1}\Psi_2,\quad
1\leq m\leq 2N,\\
\frac{\partial
F_m}{\partial\Phi_3}|_{\Phi_1=\Phi_2=0}=-\frac{1}{2}\sum_{i=1}^{m-1}
(<\Lambda^{m-i-1}\Psi_1,
\Phi_3>\Lambda^{i-1}\Psi_2-<\Lambda^{i-1}\Psi_2,
\Phi_3>\Lambda^{m-i-1}\Psi_1),\quad 1\leq m\leq 2N.
\end{array}\right.\end{eqnarray*}
When m=1, the sum terms take zero value. Firstly, after choosing
$\Phi_1=\Phi_2=\Psi_1=\Psi_2=0$ in Eq. (\ref{17}), we have
$$(0, \cdots, 0, 0, \cdots, 0, -\alpha_1\psi_{31}, \cdots, -\alpha_N\psi_{3N})^{T}=0,$$
which means that $\alpha_k, 1\leq k\leq N$ equal to zero. So eq.(\ref{17}) becomes
\begin{equation}\label{18}
\sum_{m=1}^{2N}\beta_m((\frac{\partial F_m}{\partial\Phi_1})^{T},
(\frac{\partial F_m}{\partial\Phi_2})^{T}, (\frac{\partial
F_m}{\partial\Phi_3})^{T})=0.\end{equation} Secondly, set
$\Phi_3=0$ in (\ref{18}), then we have
\begin{equation}\label{40}
\sum_{m=1}^{2N}\beta_m\lambda_j^{m-1}=0.
\end{equation}
Lastly, setting $\Psi_1=\Psi_2$ in (\ref{18}) gives
\begin{equation}\label{41}
\sum_{m=2}^{2N}(m-1)\beta_m\lambda_j^{m-2}=0.\end{equation}

Note that the determinant of the coefficient matrix  of $\beta_m$ in
Eqs. (\ref{40}) and (\ref{41}) is
$$\left|\begin{array}{ccccc}
1&\lambda_1&\lambda_1^{2}&\cdots&\lambda_1^{2N-1}\\
\vdots&\vdots&\vdots&&\vdots\\
1&\lambda_N&\lambda_N^{2}&\cdots&\lambda_N^{2N-1}\\
0&1&2\lambda_1&\cdots&(2N-1)\lambda_1^{2N-2}\\
\vdots&\vdots&\vdots&&\vdots\\
0&1&2\lambda_N&\cdots&(2N-1)\lambda_N^{2N-2}\end{array}\right|=(-1)^{N(N-1)/2}
\Pi_{1\leq i<j\leq N}(\lambda_i-\lambda_j)^{4},$$ which is not zero
when $\lambda_j, 1\leq j\leq N$ are distinct\cite{ma4}. So we have
$\beta_m, 1\leq m\leq 2N$ are zero, and then obtain that
$\{\alpha_k\}_{k=1}^{N}, \{\beta_m\}_{m=1}^{2N}$ are zero at the
same time, which contradicts to the supposition at the beginning of
the proof. Therefore, the functions $\{f_k\}_{k=1}^{N}$,
$\{F_m\}_{m=1}^{2N}$ may be functionally independent at least on
certain region of supersymmetry manifold $\mathbb{R}^{4N|2N}$. This
is the end of the proof. $\Box$

Taking into account  the preceding Hamiltonian forms and the independence of integrals of motion, we reach the integrable
property of the super FDIH with odd functions.
\begin{Theorem} The constrained systems (\ref{16}) and (\ref{34}) are completely integrable
Hamiltonian systems in the Liouville sense.
\end{Theorem}

\section{Conclusions and Discussions}

In this paper we have extended  the binary nonlinearization approach
of the (1+1)-dimensional integrable system to the super AKNS system.
By this approach, we obtained  a super FDIH with odd functions in
supersymmetry manifold $\mathbb{R}^{4N|2N}$, and their Hamiltonian
forms and integrals of motion are constructed explicitly.  The
integrable property in the Liouville sense is proved. By comparing
with the classical counterpart, the main difficulties of the
nonlinearization of the super AKNS system  are to find the
constraint in eq.({\ref{22}}) and to prove the independence of
integrals of motion in Proposition \ref{propind} due to the
appearance of odd functions. In particular, one advantage of the
nonlineariation of the super system is to construct FDIH with odd
functions in a systematic way. We know there are very few expamples
of the FDIH with odd functions \cite{lmv}.

It is a crucial fact for us to establish the nonlinearization  for
the super AKNS system  that matrix $M$ in the its problem (\ref{1})
belongs to the Lie superalgebra $B(0,1)$. However, we are not able
to do this  for the fully supersymmetric AKNS system \cite{pop}
because its spectral matrix can not be described by a certain Lie
superalgebra. For the further research related to this topic, the
nonlinearization of  the super Dirac system will be discussed in a
separate paper. Moreover, how to make nonlinarization of the fully
susy AKNS system is an interesting problem.

{\bf Acknowledgments} {\small This work is supported by the NSF of
China under Grant No. 10671187, and SRFDP of China. J. H. thanks
Professors Li Yishen(USTC,China) and Wen-Xiu Ma(USF, USA) for
valuable discussions on this topic. We are grateful to two anonymous
referees for their valuable suggestions. }


\small \baselineskip 13pt
\begin{thebibliography}{99} 
\bibitem{cao1}C. W. Cao, Sci. China A 33 (1990) 528.
\bibitem{cao2} C. W. Cao, X. G. Geng, J. Math. Phys. 32 (1991) 2323.
\bibitem{zhourg1}R. G. Zhou, J. Math. Phys. 48 (2007) 013510.
\bibitem{ma1}W. X. Ma,  W, Strampp, Phys. Lett. A 185 (1994) 277.
\bibitem{ma2}W. X. Ma,  Physica A 219 (1995)  467.

\bibitem{kon}B. Konopelchenko, J. Sidorenko, W.  Strampp, Phys. Lett. A 157 (1991) 17.
\bibitem{cl1}Y. Cheng, Y.S.Li, Phys. Lett. A 157 (1991) 22.
\bibitem{cy1}Y. Cheng, J. Math. Phys. {\bf 33} (1992) 3774.
\bibitem{os1}W. Oevel, W. Strampp, Comm. Math. Phys. 157 (1993) 51.
\bibitem{cy2}Y. Cheng, Commun. Math. Phys. {\bf 171} (1995) 661.
\bibitem{cKP2} L. L. Chau, J. C. Shaw, M. H. Tu, J. Math. Phys. 38 (1997) 4128.
\bibitem{anp1}H. Aratyn, E. Nissimov, S. Pacheva, Comm. Math. Phys. 193
(1998) 493.
\bibitem{hlc1} J. S. He, Y. S. Li, Y. Cheng, J. Math. Phys. 44 (2003) 3928.
\bibitem{pop}Z. Popowicz, J. Phys. A 23 (1990) 1127.
\bibitem{kup}B. A. Kupershmidt, Phys. Lett. A 102 (1984) 213.
\bibitem{mo}M. Gurses, O. Oguz, Phys. Lett. A 108 (1985) 437.
\bibitem{Lz1+2}Y. S. Li, L. N. Zhang, Nuovo Cimento A 93 (1986) 175;
J. Phys. A 21 (1988) 1549.
\bibitem{lz3}Y. S. Li, L. N. Zhang, J. Math. Phys. 31 (1990) 470.
\bibitem{mr} Y. Manin, I. Radul, Commun. Math. Phys. 98 (1985) 65.
\bibitem{lm} Q. P. Liu, Manuel Manas, Phys. Lett. B 485 (2000) 293.

\bibitem{ad}H. Aratyn, A. Das, Modern Phys. Lett. A 13 (1998) 1185.
\bibitem{ts1} J. C. Shaw, M. H. Tu, J. Phys. A 31 (1998) 6517.
\bibitem{anp1}H. Aratyn, E. Nissimov, S. Pacheva, J. Math. Phys. 40
(1999) 2922.
\bibitem{anp2}H. Aratyn, E. Nissimov, S. Pacheva,
Berezinian Construction of Super-Solitons in Supersymmetric
Constrained KP Hierarchies, published in{\sl
  Topcis in Theoretical Phys. vol. II} (1998),
Festschrift for A.H. Zimerman, IFT-S\~{a}o Paulo, SP-1998, pgs.
17-24(also see arXiv:solv-int/9808004)

\bibitem{ts2} M. H. Tu, J. C. Shaw, J. Math. Phys. 40 (1999) 3021.
\bibitem{cartier}P. Cartier, C. DeWitt-Morette; M. Ihl;
C. S$\ddot{a}$mann, World Sci. Publ., River Edge, NJ, (2002) 412.
\bibitem{Hu} X. B. Hu, J. Phys. A: Math. Gen. 30 (1997) 619.
\bibitem{ma3} W. X. Ma, B. Fuchssteiner, W. Oevel, Phys. A 233
(1996) 331.
\bibitem{ma4} W. X. Ma, American Mathematical Monthly, 104 (1997)
566.
\bibitem{lmv}G. Landi, G. Marmo, G. Vilasi, J. Phys. A 25 (1992) 4413.
\end{thebibliography}
\end{document}